# Preliminary Planck constant measurements via UME oscillating - magnet Kibble balance

**H Ahmedov, N Babayiğit Aşkın, B Korutlu, R Orhan**

TÜBİTAK Ulusal Metroloji Enstitüsü (UME) P.K. 54 41470 Kocaeli, Turkey

E-mail: haji.ahmadov@tubitak.gov.tr

**Abstract**

The UME Kibble balance project was initiated at the second half of 2014. During this period we have studied the theoretical aspects of Kibble balances in which an oscillating magnet generates AC Faraday's voltage in a stationary coil and constructed a trial version to implement this idea. The remarkable feature of this approach is that it can establish the link between the Planck constant and a macroscopic mass by one single experiment in the most natural way. Weak dependences on variations of environmental and experimental conditions, small sizes and other useful features offered by this novel approach reduce the complexity of the experimental setup. This paper describes the principles of oscillating magnet Kibble balance and gives details of the preliminary Planck constant measurements. The value of the Planck constant determined with our apparatus is $\boldsymbol{h/h_{90} = 1.000004}$ with a relative standard uncertainty of 6 ppm.

## 1. Introduction

In 1975, Bryan Kibble described the principles of the first moving-coil Kibble balance [1, 2], a mechanical apparatus with two measurements phases, which lays a bridge between the electrical and mechanical units in the International System of Units (SI). The Kibble balance principle together with the two macroscopic electrical quantum effects: the Josephson effect and the quantum Hall effect [3, 4] establish a link between the macroscopic mass and the Planck constant, the fundamental constant of the microworld [5]. This link provides a route for the redefinition of the kilogram [6, 7], the last base unit in SI, which is still defined by a man-made object, the international prototype of the kilogram (IPK).

Significant efforts have been devoted to construct a variety of Kibble balances in different National Metrology Institutes (NMIs) (see [8] and references therein). Most of the existing moving-coil Kibble balance experiments operate in two-phases, weighing and moving, as was originally described by Dr. Kibble, and differ in the way that the coil is moved and guided during the dynamical phase [9-20] (See also the review papers [5, 7, 21-23]). The successive measurement modes of the experiment constrain the system on testing Ampere's force law and Faraday's law of induction, simultaneously, and hence evoke the need to quantify variations in the environmental and experimental conditions between the two phases at the level of parts per billion (ppb) which complicates the experiment. High sensitivity to the changes in ground vibrations and temperature, non-linear magnetic effects and alignment issues are some examples of such complications [24-26]. In contrast to the conventional two-phase Kibble balance experiments, the BIPM watt balance is less sensitive to the changes in environmental and experimental conditions as it operates only in dynamical mode, where the coil moves with a constant velocity, allowing a simultaneous measurement scheme [27-29]. In addition to Kibble balance experiments, NIM develops Joule balance experiment with a static coil and a moving magnet, which contributes significantly to the total uncertainty due to the impact of external magnetic field [30]. A technique for implementing new generation of simplified moving coil Kibble balance have been suggested by NPL [31].

In this paper, we present the theory, basic design and preliminary results of a new generation Kibble balance, the UME oscillating magnet Kibble balance, where the coil is kept stationary but the magnet undergoes an oscillatory motion [32]. The striking feature of the system manifests itself in separation of induced and resistive voltages on the coil. As the former is an oscillating one while the latter is a constant voltage, separation of the two is achieved readily, which then warrants the simultaneous testing of the Faraday's law of induction and the Ampere's force law. Yet another prominent trait of the system lies under the adopted measurement procedure such that continuous averaging over the magnet oscillation half-cycles provides an effective mechanism for the suppression of variations in magnetic field and temperature which in turn enables the construction of both the magnetic circuit and the apparatus in smaller sizes.

The feasibility of this novel approach has been tested on the trial version of the UME oscillating magnet Kibble balance where a full-range balance is integrated into the system for practical reasons. There is an on-going measurements on the second version which operates with the Mettler Toledo AX5006 mass comparator with a resolution of 1 μg, instead.

The paper is organized as follows. Section 2 is devoted to the discussion of the UME Kibble balance apparatus. In Section 3, we outline the principles of Kibble balance. In Section 4, we present the measurement procedure for the determination of Planck constant. Section 5 summarizes the results of our measurements. We conclude in Section 6.

## 2. Description of the apparatus

The general view of the UME oscillating magnet Kibble balance is shown in Figure 1. The apparatus is placed on a concrete block separated from the foundation of the building and is enclosed inside a cabin which prevents the impact of air flow. Below, we describe the mechanical apparatus and devices used for optical and mechanical measurements.

The force measuring system consists of a full-range balance Mettler Toledo PR 10003 and a weighing pan designed for inserting a reference mass. Suspended from the balance is a stationary coil immersed in the magnetic circuit. The coil is connected to its support frame via three non-magnetic rods which are equally spaced around the circumference of the coil former.

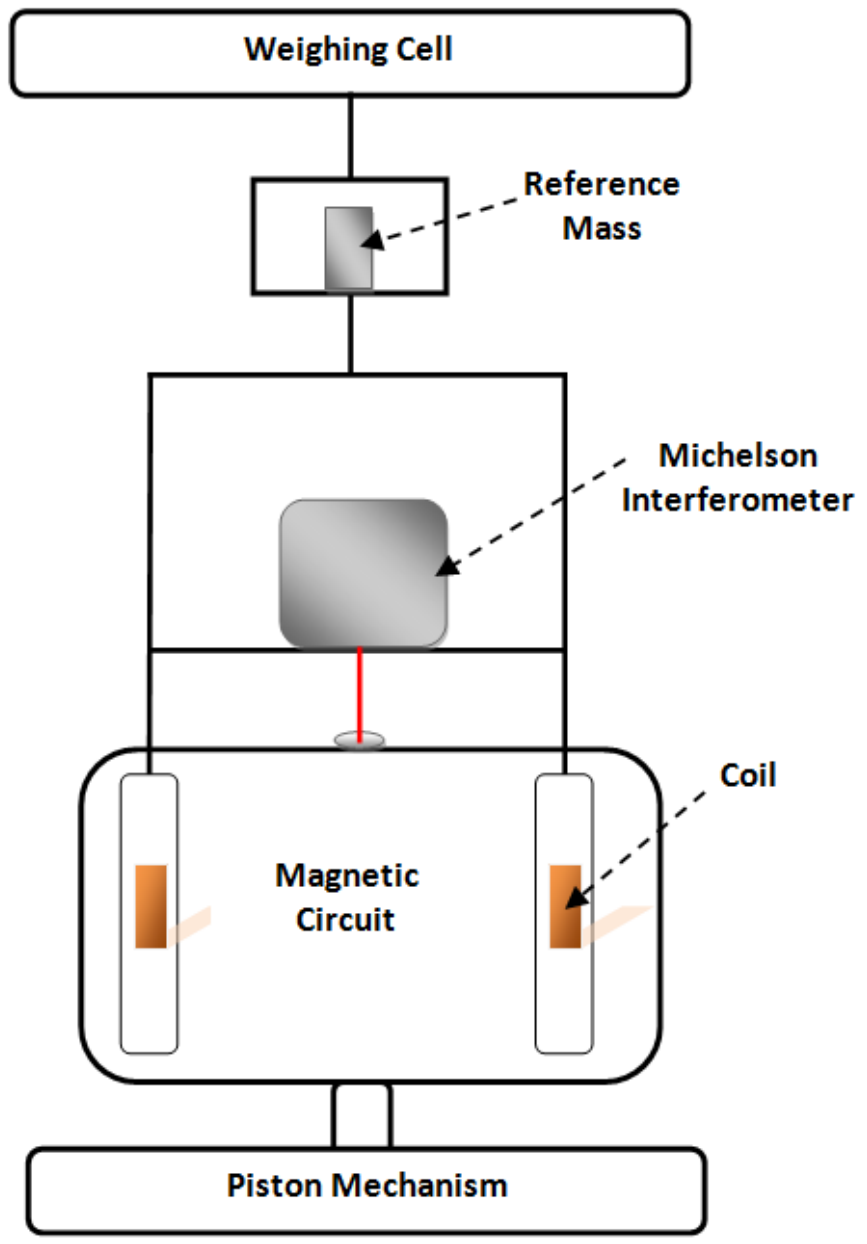

**Figure 1.** General view of the UME oscillating magnet Kibble balance apparatus.

The cross sectional view of the aforementioned magnetic circuit is given in Figure 2.

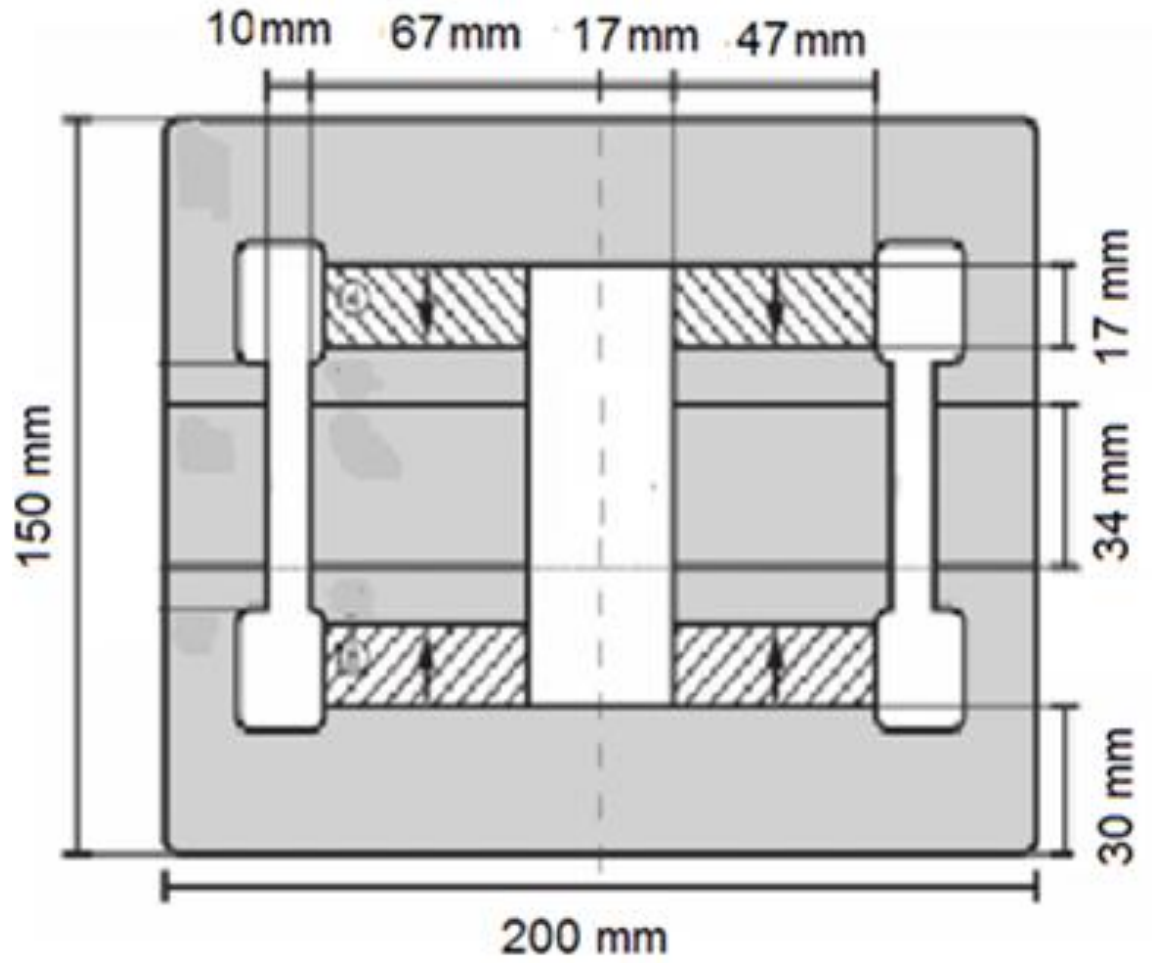

**Figure 2.** Cross-sectional view of the UME magnetic circuit. The yoke is made from iron and the permanent magnets from SmCo.

A closed type, radially symmetric magnetic circuit is used as it was confirmed to be the most practical solution for a Kibble balance realizing the kilogram [33-35]. Such a magnetic circuit is known to produce rotational and up-down symmetric fields in the air gap. The configuration shown here is a variation of the NIST4 magnet system [33] with one notable difference: The sizes in all spatial directions are reduced three times. As the magnetic field remains invariant under the rescaling, the UME magnetic circuit may be described based on the results obtained for the NIST4 magnet which yields a radial field of about $B = 0.55\,T$ in the centre of air gap.

A piston mechanism is used to provide the vertical movements of the magnet. The motion of the piston is controlled by a servo-motor and a reducer as shown in Figure 3. As the reducer decreases the output frequency of the rotating motor by a factor specific to the type of the reducer used, the oscillation frequency of the magnet and the rotation frequency of the motor will be different from each other. Since the measurement procedure in Planck constant determination picks the frequencies close to the fundemantal oscillation frequency, the effect of higher frequencies caused by motor rotation are damped. This grants the elimination of the noise created by the rotating motor. In addition, reducer decreases the torque on the motor which in turn produces a better reproducibility in the oscillatory motion. Mechanical constraints are used to minimize the angular and horizontal motions of the magnet. This is an essential issue for the reduction of misalignment uncertainties.

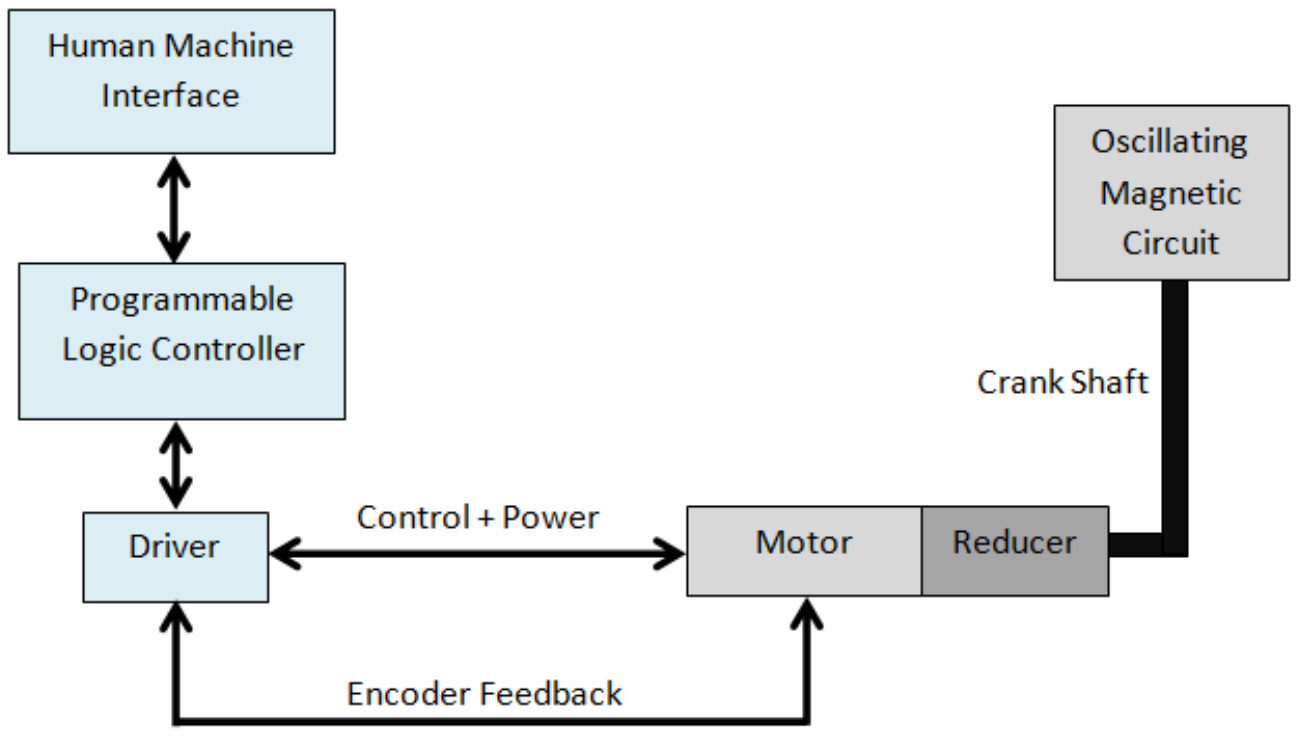

**Figure 3:** The diagram of moving system.

SIOS AE SP 2000E Michelson Interferometer with plane mirror reflector together with a tilt mechanism is placed on the surface of the coil support frame which is used for measuring the relative velocity of the coil with respect to the oscillating magnet. The laser beam is directed from a reflective mirror placed on the center of the magnet upper surface. As the platform of interferometer is rigidly attached on the coil support frame, one single interferometer seems to be sufficient for measuring the Planck constant.

## 3. Kibble balance principle

Consider a physical system, consisting of a magnetic circuit moving along the direction of the gravitational acceleration and a stationary coil, carrying an electrical current $J$, immersed in the air gap of the magnetic circuit. According to Ampere's force law, the magnetic field generated by the magnetic circuit induces a Lorentz force

$$F(t) = G(t)J(t), \tag{1}$$

in the current carrying coil. Here $G(t)$ is a geometrical factor which depends on the structure of the magnetic field and the geometry of the coil. According to Faraday's law of induction, the moving magnetic circuit induces a Faraday's voltage

$$V(t) = G(t)u(t), \tag{2}$$

across the ends of the coil. Here $u(t)$ is the velocity of the coil with respect to the magnetic circuit (in the rest of the paper we will use the term *relative coil velocity* instead of writing each time *the velocity of the coil with respect to the magnetic circuit)*. By combining the Ampere's force law and the Faraday's law of induction we arrive at the Kibble balance principle

$$\frac{h}{h_{90}} V(t)J(t) = F(t)u(t). \tag{3}$$

The ratio $h/h_{90}$ ($h$ is the Planck constant and $h_{90}$ is the conventional value of the Planck constant) in the left hand side of Eq. (3) appears due to the fact that the electrical units of voltage and resistance have been measured based on the conventional values of the Josephson and the Klitzing constants since 1990 [36].

## 4. The Planck constant measurement procedure

The measurement procedure of the UME oscillating magnet Kibble balance is described below. Magnetic circuit commits a nearly periodic motion in the vertical direction with a fundamental period $T$. Figure 4 illustrates the oscillatory motion of the magnet. The data is obtained via the Michelson Interferometer.

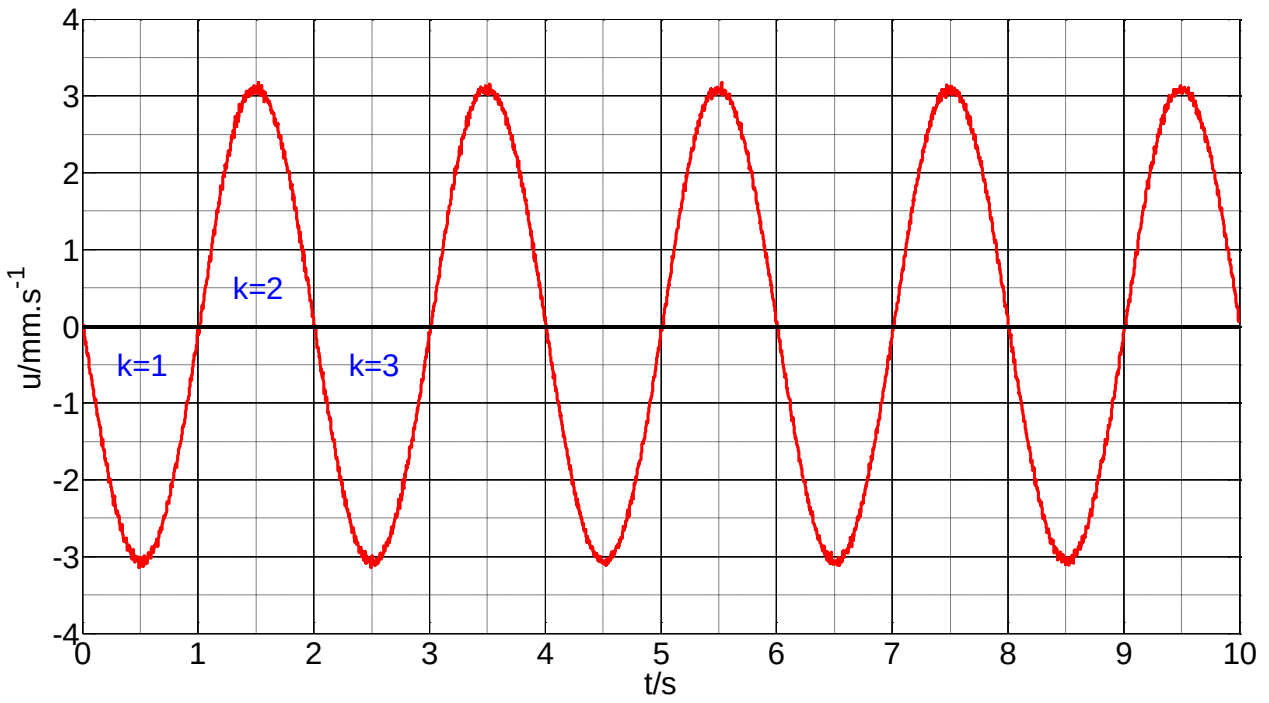

**Figure 4.** The relative coil velocity. The half-cycles are indicated by assigned $k$ values.

In the measurement procedure, the separation of the statical (mean value) and dynamical (deviation from the mean value) properties of the physical quantities is proven to be useful. For an arbitrary physical quantity $g$, we define the deviation of this quantity from its mean as (in the rest of the paper, we use the term *deviation* instead of writing each time *the deviation from the mean*)

$$\Delta g = g - \langle g \rangle, \tag{4}$$

where the mean value $\langle g \rangle$ is the average over the integration time, $\tau$. The integration time $\tau = NT$ of the Kibble balance experiment is chosen to be a multiple of the fundamental period $T$. The number $N$ is determined in a statistically optimal manner. We divide the integration time, $\tau$ into $2N$ half-cycles and define the initial conditions in such a way that the deviation of the velocity $\Delta u$ is close to zero at $t_k = \frac{T}{2}(k-1)$ for $k = 1, 2, \ldots, 2N$. The average of the physical quantity $g$ over the $k^{th}$ half-cycle $(t_k, t_{k+1})$ is determined by

$$\langle g \rangle_k = \frac{2}{T} \int_{t_k}^{t_{k+1}} g(t)\, dt. \tag{5}$$

We define the average of $g$ over the $2N$ half-cycles of the velocity $u$ by means of the formula

$$\{g \mid u\} = \frac{1}{2N} \sum_{k=1}^{2N} \frac{\langle \Delta g \rangle_k}{\langle \Delta u \rangle_k}. \tag{6}$$

The averaging process described in Eq. (6) can also be done over the $2N$ half-cycles of the Faraday voltage $V$ as $\{g \mid V\}$ instead of $\{g \mid u\}$, because their minima coincide with each other. Throughout the text we use the definitions in Eqs. (4), (5) and (6) for different physical quantities.

In accordance with the definitions given above, one may write the fundamental formula for Planck constant as to be observed in the UME oscillating-magnet Kibble balance apparatus as

$$\frac{h}{h_{90}} = \left\{ \frac{F}{J} u \,\middle|\, V \right\}, \tag{7}$$

where the Lorentz force $(F)$ and relative coil velocity $(u)$ along the direction of gravitational acceleration, the Faraday voltage $(V)$ in the coil and the current $(J)$ flowing through the coil are the measured quantities. For practical reasons, using the Ampere's force law given in Eq. (1), we rewrite the fundamental equation for Planck constant as

$$\frac{h}{h_{90}} = \frac{\langle F \rangle}{\langle J \rangle} Q \, \{u \mid V\}. \tag{8}$$

where, the factor

$$Q = 1 + \left\{ \frac{\Delta G}{\langle G \rangle} u \,\middle|\, V \right\} / \{u \mid V\}, \tag{9}$$

includes the effects of inhomogenities in the air gap of the magnetic circuit, imperfectness of the coil, thermal effects and other non-linear effects as described in the Subsection 4.1.

### *4.1 Geometrical Factor*

The mean deviation of the geometrical factor up to linear order approximation may be represented as follows

$$\Delta G = \langle G \rangle (\Delta f + \Delta \zeta), \tag{10}$$

where $f$ and $\zeta$ describe the vertical inhomogenities of the magnetic field in the air gap and the temperature effects, respectively.

The vertical inhomogenities in the air gap of the magnetic circuit are determined by static force measurements in different positions of the magnet and may be modeled as an $n^{th}$ degree polynomial [10, 12] in the following form

$$f = \sum_n \Pi_n \left(\frac{z - z_0}{a}\right)^n, \qquad (11)$$

where $a = \sqrt{2\langle(\Delta z)^2\rangle}$ is the oscillation amplitude of the magnetic circuit, $z$ is the instantaneous vertical position and $z_0$ is the initial position of the magnetic circuit. Note that, it is the displacement measured by the interferometer not the position. Therefore, $z - z_0$ is the measured quantity. The coefficients $\Pi_n$ describe the inhomegenities in the $n^{th}$ order. The unknowns $z_0$ and $\Pi_n$ are determined by using the least square estimation in $G$ based on static force measurements in different positions of the magnet.

The effect of changes in the temperature of the permanent magnet and the yoke on the magnetic field are described by

$$\Delta\zeta = \alpha\, \Delta T, \qquad (12)$$

where $\Delta T$ is mean deviation of the permanent magnet temperature and $\alpha = -3 \times 10^{-4}\, K^{-1}$ is the temperature parameter of the SmCo magnets. Since temperature variations are slow physical processes, the spectral density of the temperature concentrates near the zero frequency. The averaging procedure over $2N$ half-cycles, on the other hand, picks up harmonics of the function $\Delta\zeta$ with frequencies which are multiples of the oscillation fundamental frequency. Therefore, the averaging process strongly suppresses the thermal effects on the magnetic field.

Additional non-linear effects such as external magnetic field and eddy currents are not included in the scope of this paper as the present measurement uncertainty is higher than the expected uncertainties by these effects.

*4.2 Faraday Voltage*

Figure 5 illustrates the equivalent electrical circuit of the oscillating magnet Kibble balance. Note that the output voltage $W$ of the coil is not the same as the Faraday's voltage $V$ of the coil such that

$$V = W + LC\,\ddot{W} + RC\dot{W} - IR - L\dot{I}. \qquad (13)$$

Likewise, the current $J$ flowing through the coil is not equal to the measured current $I$ on the standard resistance $R_{st}$ but

$$\mathrm{J} = \mathrm{I} - \mathrm{C}\dot{\mathrm{W}}. \qquad (14)$$

The measured electrical quantities of the UME Kibble balance experiment is the output voltage $W$ of the coil circuit and the output voltage $W_A = IR_{st}$ across the standard resistance.

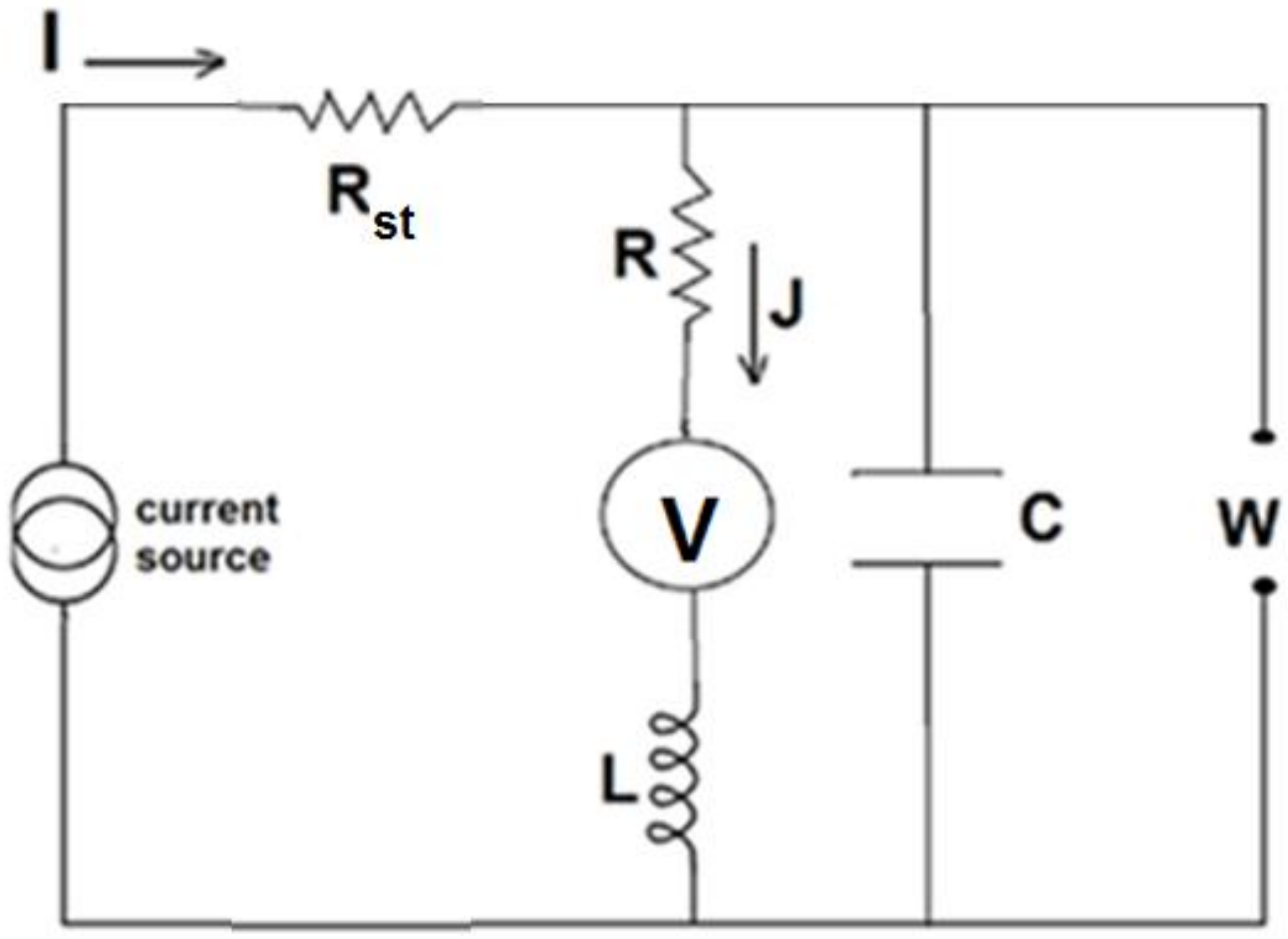

**Figure 5.** Equivalent circuit of the oscillating magnet Kibble balance.

Additional measured quantities are the capacitance $C$ and the inductance $L$ of the coil in the magnetic circuit. Using Eq. (13), we observe that the only unmeasured term in the mean deviation $\Delta$V of the Faraday's voltage is the temperature variations of the coil resistance of the form

$$\Delta R = \beta\langle R\rangle\Delta T, \qquad (15)$$

where $\beta = 4 \times 10^{-3}\, K^{-1}$ is the temperature coefficient. Similar arguments with the thermal effect on magnetic field $\Delta\zeta$, lead us to conclude that the uncertainties arising from the temperature changes in the coil resistance decreases with the oscillation frequency.

*4.3 Alignment*

The Kibble balance principle in Eq. (3) is valid when the relative velocity of the coil is directed along the vertical (gravitational) axis. In practice, however, the magnetic circuit may also have horizontal and angular motions. Thus, a misalignment term of the form $\vec{F} \cdot \vec{u} + \vec{K} \cdot \vec{\omega}$ should be added to the right hand side of Eq. (3). Here $\vec{F}$ is the horizontal component of the Lorentz force, $\vec{u}$ is the horizontal velocity of the coil with respect to the magnet such that $\vec{u} = \vec{u}_c - \vec{u}_m$, where $\vec{u}_c$ is the horizontal coil velocity and $\vec{u}_m$ is the horizontal magnet velocity, $\vec{\omega}$ is the angular velocity of the coil with respect to the magnetic circuit and $\vec{K}$ is the torque in the coil measured with respect to the point at which the laser beam hits the magnet. Such misalignment effects reflect themselves as uncertainty on Planck constant measurement in the following form

$$\frac{\delta h^{(a)}}{h_{90}} = \frac{\langle \vec{F} \rangle}{\langle F \rangle} \cdot \{\vec{u} \mid u\} + \frac{\langle \vec{K} \rangle}{\langle F \rangle} \cdot \{\vec{\omega} \mid u\}. \qquad (16)$$

Measurements in the UME Kibble balance apparatus indicate that under the rotation of the magnet together with the coil, while the direction of dynamical alignment parameters $\{\vec{u} \mid u\}$ and $\{\vec{\omega} \mid u\}$ remain almost the same, the direction of the static alignment parameters $\langle \vec{F} \rangle$ and $\langle \vec{K} \rangle$ change accordingly. As the misalignment uncertainty is determined by the scalar products, it may be suppressed either by averaging Planck constant values obtained at different rotation angles or by finding the orientation of the magnet where these scalar products are minimum.

## 5. Measurements and results

In UME oscillating-magnet Kibble Balance set-up, the duration of the experiment for obtaining a Planck constant value is 400 s which includes 10 sets of measurements of 30 s and data transfer of 10 s which is required due to the restriction on the memory of Keysight 3458A digital multimeter. The oscillation frequency of the experiment is 0.5 Hz and the number of half-cycles in each set is equal to 30. The sampling frequency of the PR 10003 balance is set to 20 Hz and Keysight 3458A digital multimeters are 1 kHz. The diagram below describes the data acquisition process of our measurement set-up.

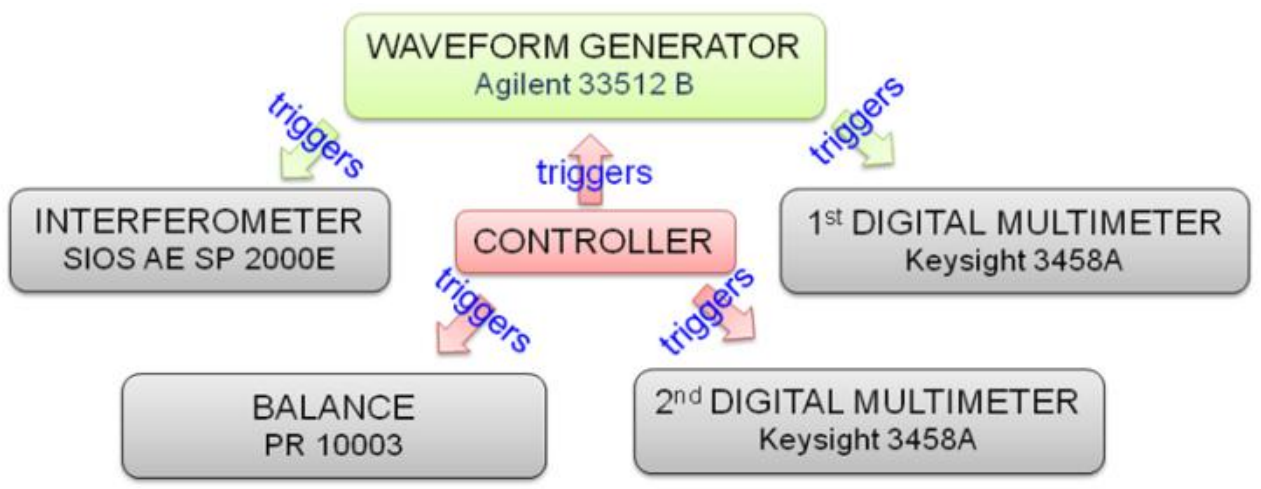

**Figure 6.** The diagram of data acquisition in UME oscillating magnet Kibble balance experiment

Although Planck constant measurements in UME oscillating magnet Kibble balance experiment are carried out in single phase, we present the results in two stages, as in conventional Kibble balance experiments, in an attempt to make the paper more readable.

### *5.1 Moving Stage*

The voltage drop W across the coil circuit is measured by Keysight 3458A digital multimeter which was calibrated by means of 10 V programmable Josephson voltage standard. The output voltage $W$ of the coil circuit is shown in Figure 7.

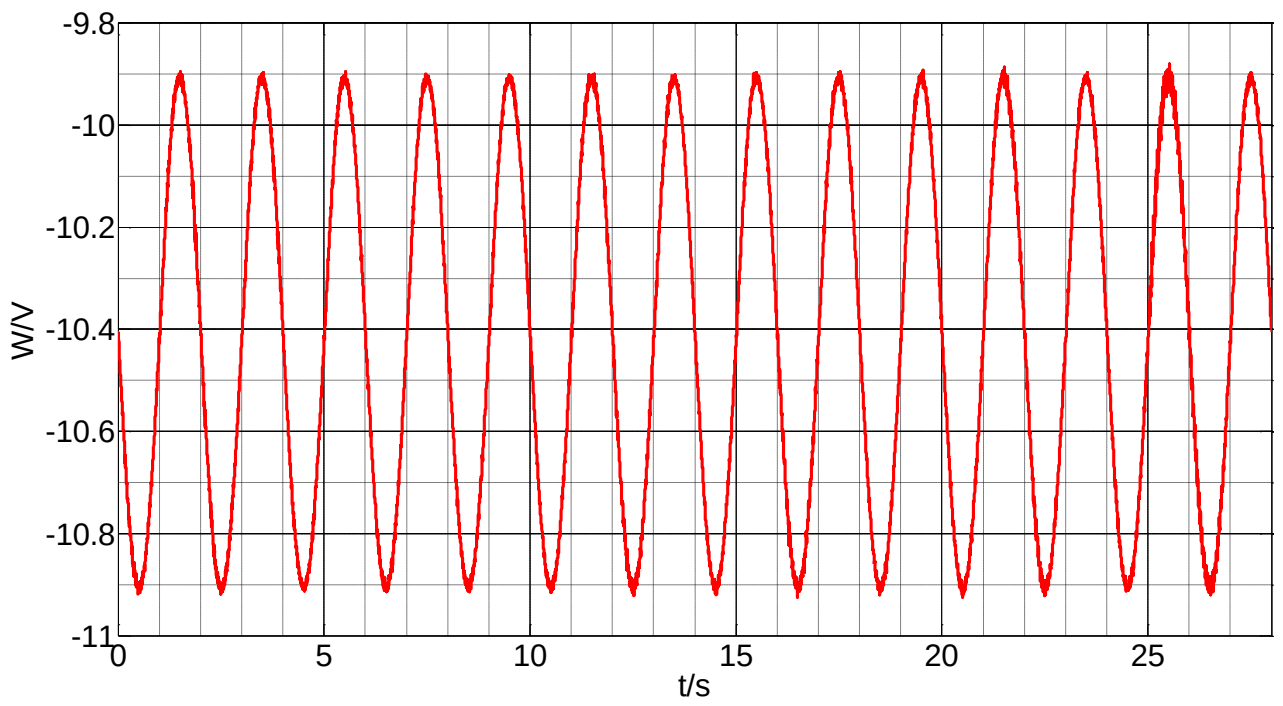

**Figure 7.** The output voltage $W$ of the coil circuit.

Michelson Interferometer with plane mirror reflector is used in measuring the relative coil velocity. The synchronization between the relative coil velocity $u$ and the Faraday's voltage $V$ across the coil is an essential issue in Planck constant measurements. The synchronization is achieved by using two channel Keysight 33512B waveform generator as trigger (See the diagram in Figure 6). Two channel option of the waveform generator allows us to adjust the phase difference between the interferometer and multimeter. FFT is used in determining the phase angle. The circuit diagram of electrical and velocity measurements is given in Figure 8.

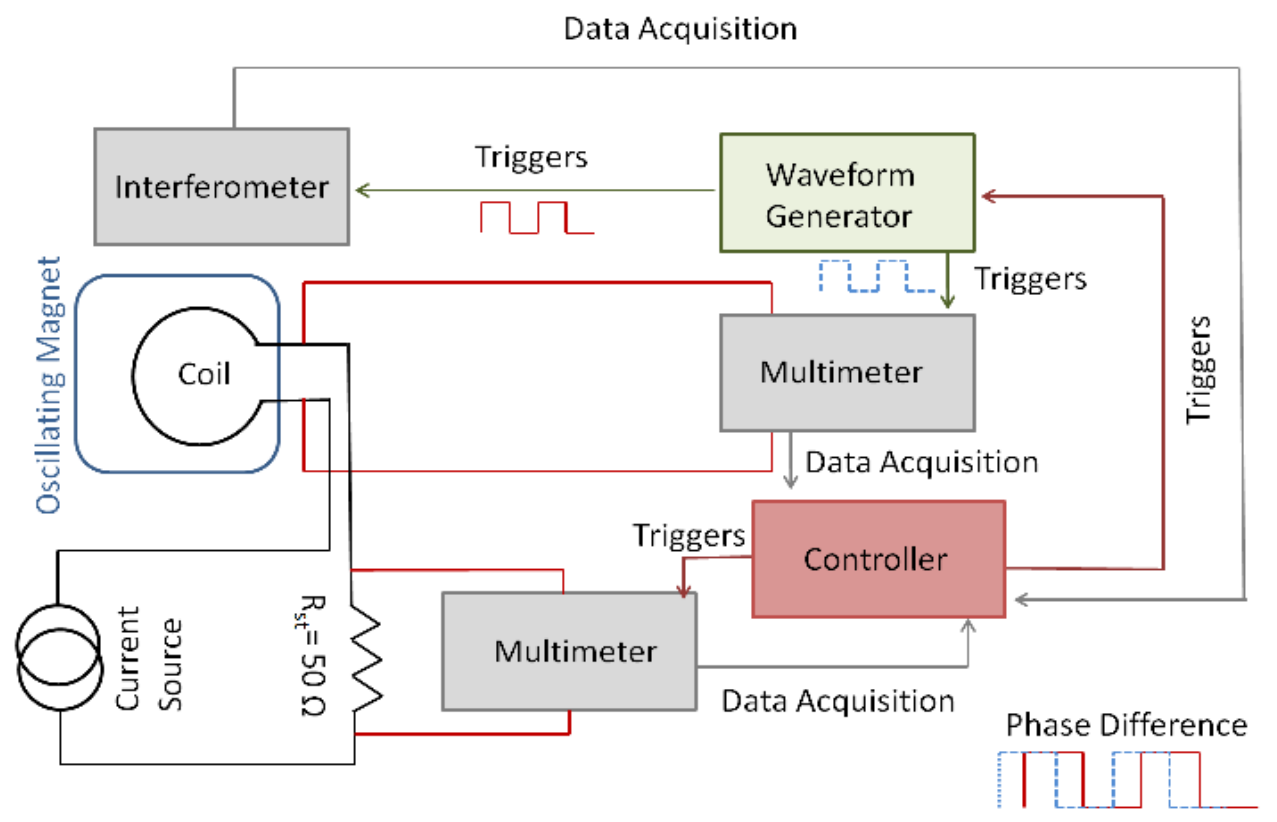

**Figure 8.** The circuit diagram of electrical and velocity measurements.

The synchronized data is represented in Figure 9 and zoomed in Figure 10.

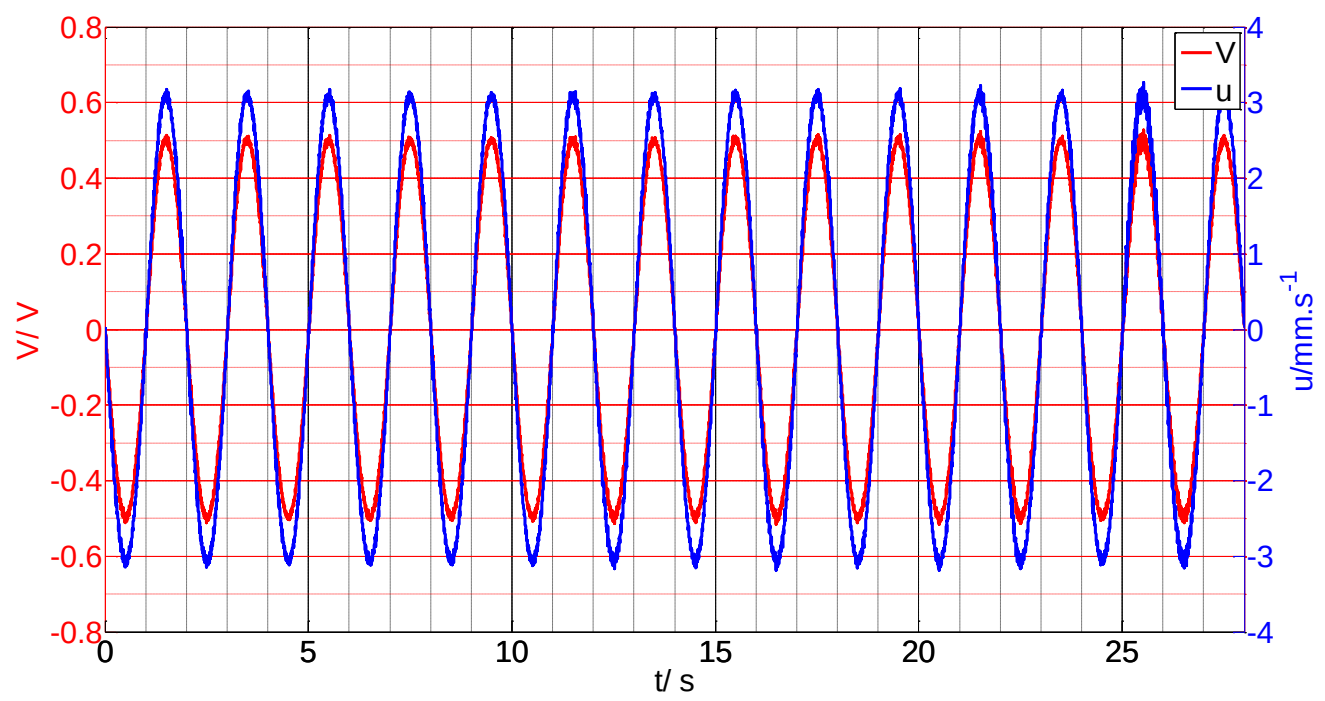

**Figure 9.** The synchronization between the Faraday Voltage $\boldsymbol{V}$ (red) and the relative coil velocity $\boldsymbol{u}$ (blue).

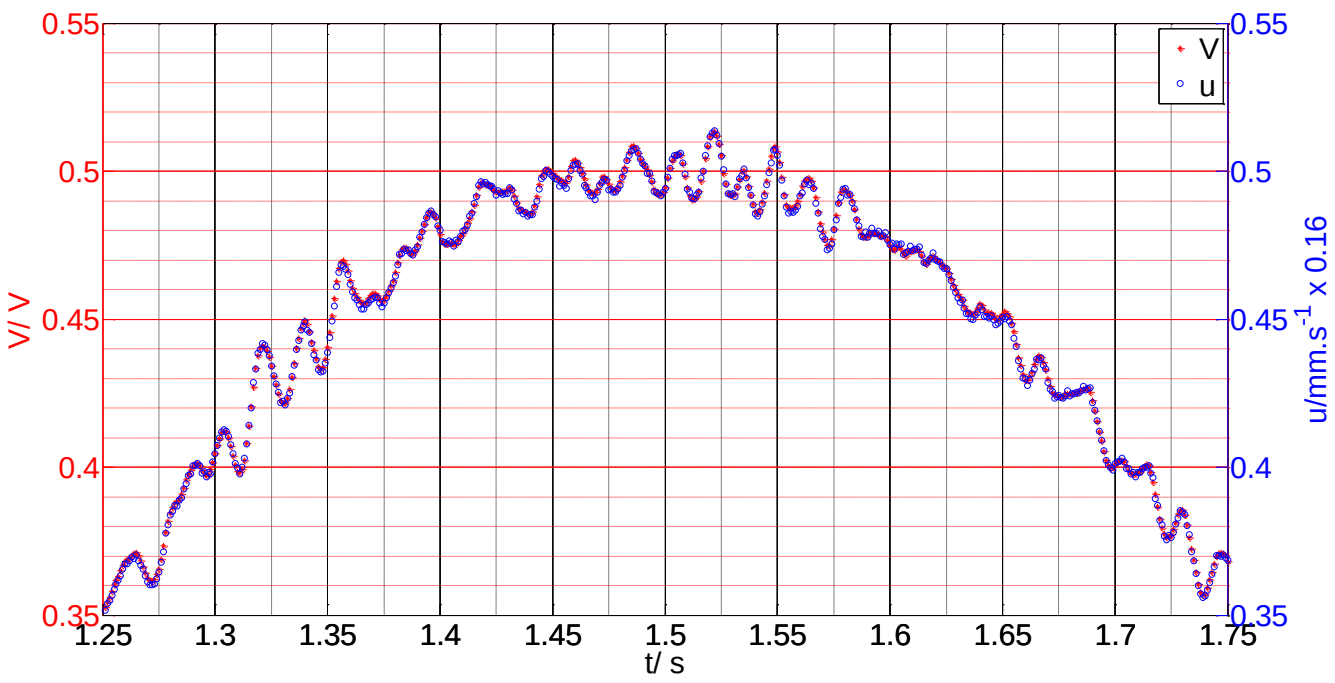

**Figure 10.** Zoom-in plot of the synchronization between the Faraday Voltage $\boldsymbol{V}$ (red stars) and the relative coil velocity $\boldsymbol{u}$ (blue circles). The velocity data is rescaled with 0.16 for a better illustration of the synchronization.

*5.2 Weighing Stage*

The electrical current measurements are carried out by using Keysight 3458A digital multimeter across the two Tinsley 5658A 100 Ω standard resistors connected in parallel. The voltage drop $W_A$ across these resistors is shown in Figure 11. The current source, the coil and the standard resistors form a closed loop as shown in Figure 5. This is why there appears an oscillating voltage across the resistors caused by the Faraday's voltage, in addition to the DC voltage supplied by the Keithley 6220 precision current source.

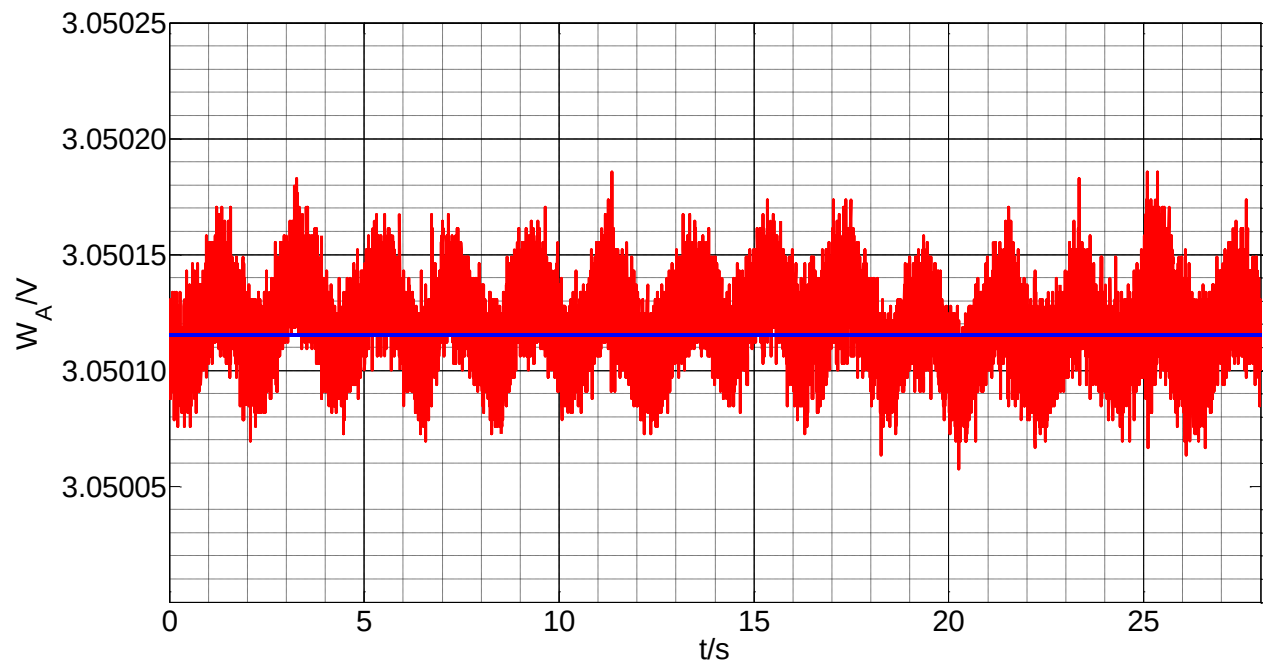

**Figure 11.** The voltage drop $\boldsymbol{W_A}$ across the standard resistors. The solid blue line represents the mean value of the data.

The inhomogenities of the geometric factor are measured by static force measurements in different positions of the magnet. The force measurements are carried by Mettler Toledo PR10003 balance. We use ten equally spaced positions in these measurements. The analysis is completed by using sixth order polynomial fitting. The plot of the polynomial is given in Figure 12. Here, $z(t)$ indicates the vertical displacement between the so called symmetric centers of the magnet and the coil. By definition, the zero displacement $\boldsymbol{z(t) = 0}$ coincides with the maximum value of the polynomial. It is important to point out that the polynomial is constructed as a function of the measured quantity $\boldsymbol{(z(t) - z_0)/a}$ not $\boldsymbol{z(t)}$ itself. We obtain the value of $z_0/a$ from the plot as the distance between the origin and the projection of the maximum value of the polynomial on the x-axis. Finally we substitute this polynomial in Eq. (9) to arrive at the $Q$ factor.

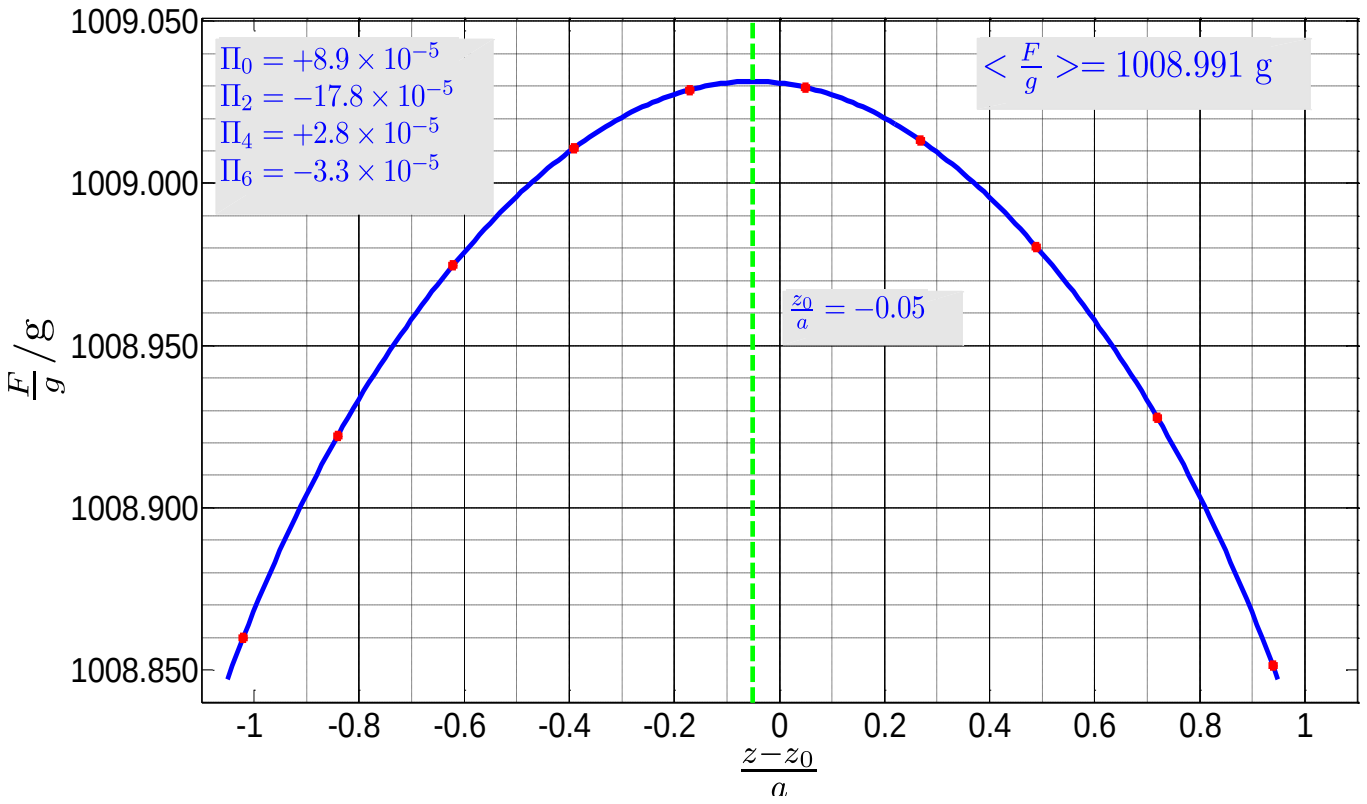

**Figure 12.** Vertical inhomogenities of the geometric factor fitted by the sixth order polynomial. The odd coefficients of the polynomial are not listed as averaging procedure over the half-cycles suppresses them and also they are order of magnitude smaller compared to the even ones.

### *5.3 Alignment Measurements*

The alignment procedure is explained extensively in [10]. These approaches are implemented in our experiment by transforming our equations in a commonly used form where we use averages of the physical quantities instead of constant values and norm of the concerned parameters rather than their components (see Table 1). As was described in Subsection 4.3, before commencement of Planck constant measurements, we find the orientation minimizing the scalar product in Eq. (16) and fix the positioning for maintaining this configuration. It is important to note that, the scalar product angles are not listed as the desirable orientation is found by set and measure sequence. The dynamical alignment parameters are measured via two Keyence Laser Displacement Sensor (LDS) heads (LK-H027). The horizontal velocity of the coil is given in Figure 13. As it could be seen from the figure, the magnet oscillation frequency does not coincide with the frequency of the coil. The latter may be one of the resonance frequencies of the apparatus excited by the oscillation of the magnet.

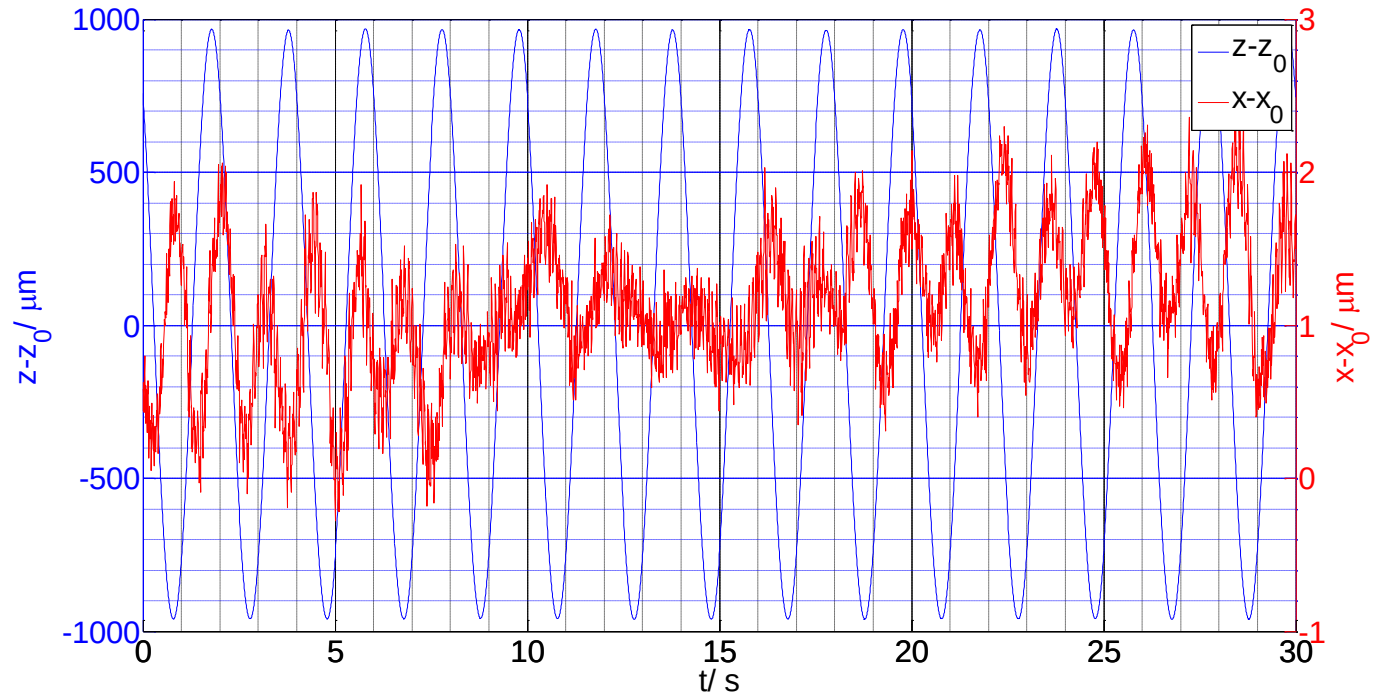

**Figure 13.** Horizontal displacement of the coil (red) and vertical displacement of the magnet with respect to the coil (blue).

The analysis of the data indicates that the norm of dynamical alignment parameters $\{\vec{u}_c|\,u\}$ in Eq. (16) is of order $10^{-5}$. By means of reasonable alignment of the horizontal force, the horizontal movements of the coil do not affect the accuracy in Planck constant measurements. The same is true for the angular motions of the coil. The only alignment parameter that would affect the Planck constant measurement comes from the magnet motion of which the numerical values are summarized in the Table 1 in the form of norm of the corresponding vectors. The torque $\vec{K}$ in the coil is obtained by using three SMD2551 single point load cells placed on the support frame of the coil at equal angles. The horizontal force $\vec{F}$ is obtained from the handler horizontal displacement data when the current is switched on.

**Table 1.** The dynamical and static alignment parameters

| Measured Quantity | Value |
|---|---|
| $\lvert\{\vec{u}\,\lvert u\}\rvert$ | 0.011 |
| $\lvert\{\vec{\omega}\lvert\, u\}\rvert$ | 0.0291 $\boldsymbol{m^{-1}}$ |
| $\lvert\langle\vec{F}\rangle\rvert/\langle F\rangle$ | 0.001 |
| $\lvert\langle\vec{K}\rangle\rvert/\langle F\rangle$ | 0.0002 m |

After finding the optimum orientation making the scalar product in Eq. (16) minimum, we arrive at the following uncertainty values as summarized in Table 2. Uncertainties negligible at this level are not included in the table.

**Table 2.** The uncertainty budget in Planck Constant determination via UME Oscillating Magnet Kibble Balance.

| Measured Quantity | Uncertainty (ppm) |
|---|---|
| Force | 4 |
| Electrical | 3 |
| Alignment | 3 |
| Planck Constant | 6 |

## 6. Conclusion

We have proposed the theory and the basic design for the oscillating magnet Kibble balance which provides a link between a macroscopic mass and the Planck constant in a most natural way. High accuracy, small sizes, ability to operate in normal laboratory conditions without any necessity to complex vibration isolation or temperature control systems and short duration of the experiment are all major advantages of the oscillating magnet approach. The trial version UME Kibble balance apparatus allows us to determine the Planck constant with a relative standard uncertainty of 6 ppm and the analysis of the long-term data indicates that the primary source of uncertainties of this apparatus is related to alignments. One may suppress the uncertainty related with the misalignments of the magnet and coil either by averaging Planck constant values obtained at different rotation angles or by finding the orientation of the magnet minimizing the scalar products (see Eq. (16)). There is an ongoing work with measurements on the second version of UME oscillating magnet Kibble Balance system.

## Acknowledgement

We would like to thank to UME management for supporting this project and our colleagues from Physics and Mechanics departments who have provided assistance for this project. We would like to thank to the anonymous referees for their valuable comments and suggestions.